\documentclass[reprint,amsmath,amssymb,
 aps,twocolumn]{revtex4-2}

\usepackage{graphicx}

\usepackage{physics}
\usepackage{hyperref}
\usepackage{xcolor}

\usepackage{comment}

\begin{document}

\title{\textit{Hello Quantum World!}\\ A rigorous but accessible first-year university course in quantum information science}

\author{Sophia E. Economou}\email{economou@vt.edu}
\author{Edwin Barnes}
 
\affiliation{Department of Physics, Virginia Tech, Blacksburg, Virginia 24061, USA}

\begin{abstract}
Addressing workforce shortages within the Quantum Information Science and Engineering (QISE) community requires attracting and retaining students from diverse backgrounds early on in their undergraduate education. Here, we describe a course we developed called Hello Quantum World! that introduces a broad range of fundamental quantum information and computation concepts in a rigorous way but without requiring any knowledge of mathematics beyond high-school algebra nor any prior knowledge of quantum mechanics. Some of the topics covered include superposition, entanglement, quantum gates, teleportation, quantum algorithms, and quantum error correction. The course is designed for first-year undergraduate students, both those pursuing a degree in QISE and those who are seeking to be `quantum-aware'. 
\end{abstract}

\maketitle

\section{Introduction}

The enormous interest in quantum information science and engineering (QISE) from academia, industry, and governments worldwide has led to new K-12 initiatives such as the National Q-12 Education Partnership and Q2Work~\cite{q12} and has prompted many universities to start degree programs, including minors, masters, and even majors in QISE. An associated challenge is structuring these degrees and selecting the courses to provide a solid training in QISE. Considering the diverse nature of the field, these degrees can come in many different flavors and place emphasis on different aspects of QISE~\cite{undergrad_education}. 

Regardless of the specific direction they follow, the students need to learn the key concepts on which they can build. These include the idea of a qubit, of superposition, of a quantum circuit, and of entanglement. Additionally, ideally students should also learn some basics about quantum algorithms, quantum cryptography, and quantum error correction. Exposure and familiarity with this material should occur early on, both to build the foundations for the more advanced material, and also to convey to the students why this field is exciting and different from what they have seen before in their STEM classes. 

At first glance, early exposure in a substantive fashion may appear as a very challenging task, considering that this material traditionally follows semesters of advanced math and quantum mechanics courses. We have found, however, that with only high-school level math, mostly addition, multiplication, a very small amount of algebra, and a basic understanding of probability, students can learn in a rigorous and, we would argue, much more intuitive way the fundamentals of quantum information and computation. This is done through the use of a pictorial representation of qubits, gates, and quantum circuits, combined with access to IBM’s Quantum Composer~\cite{IBMcomposer}. After years of creating and refining such an approach for outreach, we designed and taught a semester-long freshman course along the same lines called Hello Quantum World! (HQW!). We received very positive feedback from the students, and we found it to be a fun course to teach. We have since been contacted by numerous colleagues from universities around the US and abroad asking for advice and help in setting up a similar course. This short paper is aimed at such colleagues and other instructors who are interested in starting a similar course at their institutions, be it at the college freshman or even high-school level.

In addition to its crucial role for building the required foundations, early exposure through an accessible formalism enables equal opportunity, piques interest, allows the students a conceptual grasp of quantum information/computation and even of the underlying linear algebra structure (without explicit teaching of linear algebra). Hopefully, such an approach also helps to increase diversity: without the need for advanced math and prerequisite courses, the difference between more and less prepared students is practically diminished. Indeed, from our experience we found that the students who had come from prestigious high schools performed on average similarly to those coming from less privileged backgrounds. We hope that in the future, a careful evaluation of our course by professionals can assess the validity of this empirical finding and identify ways to further improve on this aspect. 

\begin{figure*}[htp]
    \includegraphics[width=0.6\textwidth]{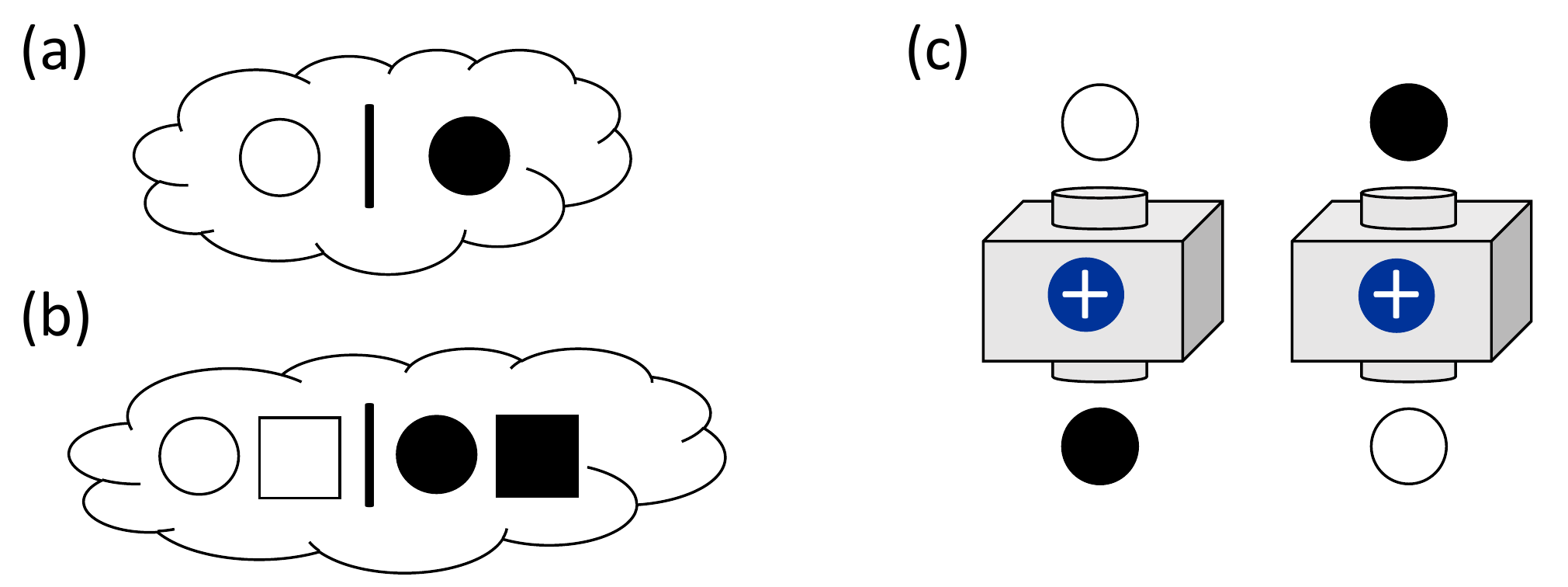}
    \caption{(a) The single-qubit state $\ket{+}$ and (b) the two-qubit Bell state $(\ket{00}+\ket{11})/\sqrt{2}$ represented as mists. (c) The single-qubit NOT gate represented as a box that flips the value of the (qu)bit input at the top.}
    \label{fig:mists_NOT}
\end{figure*}

HQW! is the centerpiece of a multidisciplinary QISE minor degree that was recently established at Virginia Tech. The minor is available to students from 7 different departments and undergraduate degree programs, including Chemistry, Computational Modeling and Data Analytics, Computer Science, Electrical and Computer Engineering, Materials Science, Mathematics, and Physics. The minor includes four mandatory courses, of which HQW! is one, while the other three are linear algebra, quantum software/programming, and an upper-level introductory course on quantum information theory. In addition, students in their final year must take either an advanced course on quantum information from the CS department or a Physics course called Quantum Optics and Qubit Processors, which introduces the central concepts and leading qubit platforms on which existing and future quantum processors are based. Beyond these five courses, students are required to take another two QISE-related courses chosen from a long list of options; students can typically satisfy this requirement by selecting courses from their home department that also count toward their major. A more detailed description of the minor, along with descriptions of other QISE degree programs, can be found in Ref.~\cite{undergrad_education}. Because HQW! is a mandatory course for the minor, we have purposely designed it to be both broadly accessible and rigorous so that it provides a firm foundation in QISE concepts in preparation for more advanced courses to come.

\section{Short description of the course}

The HQW! course is based on a pictorial approach, in which qubits are represented by shapes and logic gates by boxes. The basis states $|0\rangle$ and $|1\rangle$ are represented by colors, white and black respectively. Superpositions are denoted by clouds (or `mists’), with different configurations are separated by bars. In the first part of the course, we limit ourselves to states with real, integer coefficients (positive or negative). Different qubits are represented by different shapes. For concrete examples, see Fig.~\ref{fig:mists_NOT}, which shows the single-qubit state $|+\rangle = \frac{|0\rangle + |1\rangle}{\sqrt{2}}$, the two-qubit Bell state $\frac{|00\rangle + |11\rangle}{\sqrt{2}}$, and a single-qubit NOT (i.e., Pauli X) gate, with state $|0\rangle$ being transformed to $|1\rangle$ and vice versa. 

\begin{figure*}[htp]
    \includegraphics[width=0.9\textwidth]{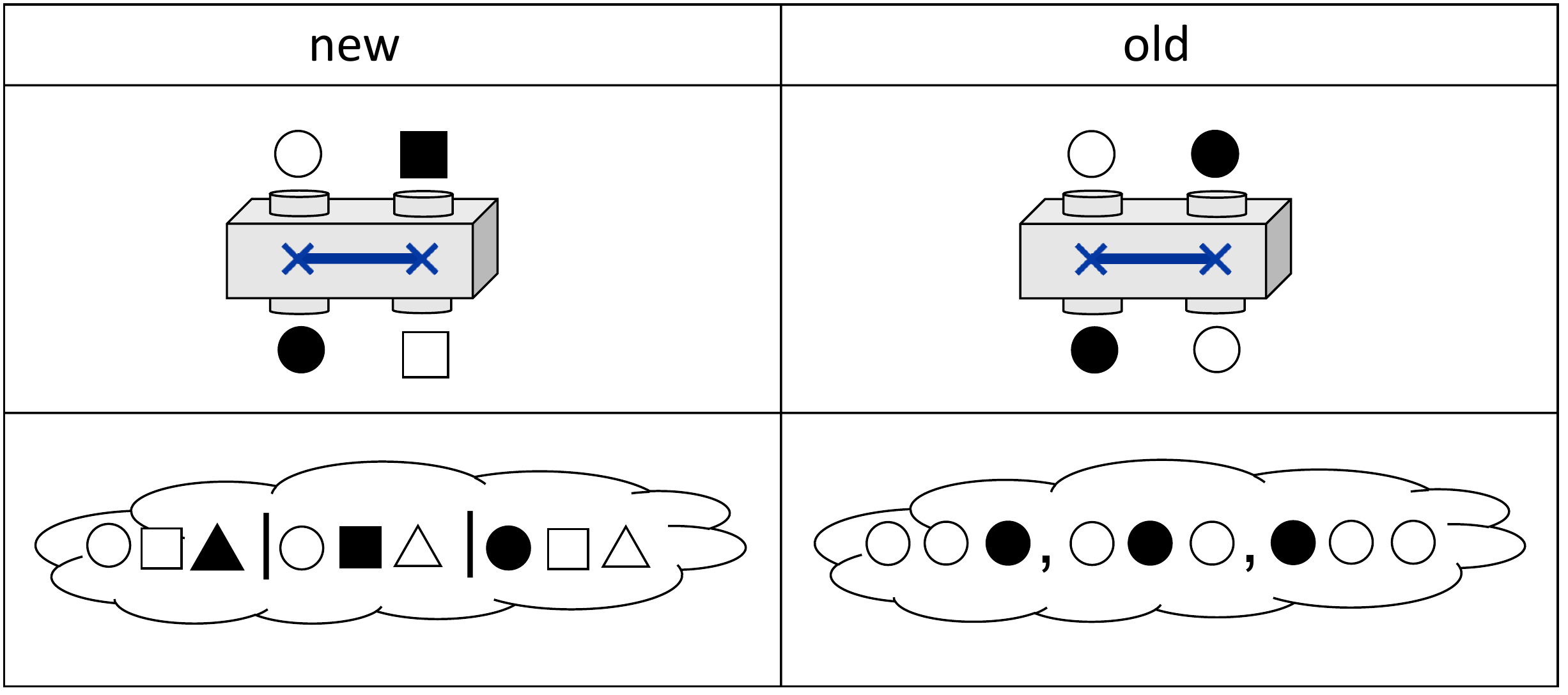}
    \caption{Examples illustrating the differences between the new and old notation in the pictorial formalism. The SWAP gate and the three-qubit W state are shown in both notations. The old version is the one introduced in Ref.~\cite{Rudolph-book}, while the new version is used in HQW!. In the new notation, different qubits are represented by different shapes, and distinct terms in a mist are separated by bars rather than commas.}
    \label{fig:notational_changes}
\end{figure*}

We note here some differences compared to an earlier paper, where we had described our outreach program~\cite{Economou_arxiv2020}. While the basic features are the same (pictorial representation combined with use of the IBM Quantum Composer), we have made key updates. The main change from our earlier approach, which was following the formalism introduced in Terry Rudolph’s popular science book Q is for Quantum~\cite{Rudolph-book}, is that different shapes are used to represent different qubits. This makes a huge difference in students' understanding and avoids a common source of confusion: by using the same shape for all qubits, it may be difficult for students to distinguish between, e.g., a superposition state of one qubit and a two-qubit state. The use of distinct shapes makes it easy to immediately see how many qubits are involved in a given state, as it coincides with the number of distinct shapes. Moreover, the use of different shapes enforces the idea of a physical system versus its state. For example, consider the SWAP gate. When a single shape is used, it is not transparent that the qubits are not swapped but their state is instead. By using distinct shapes, this becomes clear, as shown in Fig.~\ref{fig:notational_changes}. The figure also illustrates how the W state looks in the two notations. In addition to using different shapes for different qubits, in the new notation, we also separate terms in a mist using bars rather than commas. This makes the notation a little more natural when we rotate mists to orient them vertically in order to facilitate the mapping to standard circuit notation, which the students do when they work with the IBM Quantum Composer (see below).

The course begins with the introduction of the shapes as (classical) bits along with boxes representing classical gates (i.e, gates that take computational basis states to computational basis states), including NOT, CNOT, and SWAP. Quantum circuits are constructed using these gates, and students practice with classical logic. Next, quantum superpositions are introduced through the use of the cloud (or mist) representation, and the Hadamard gate is introduced as the only intrinsically quantum gate used throughout the course. Students are taught how superpositions are passed through gates, and they practice with quantum circuits. Measurement is discussed (probabilistic nature and how to calculate probabilities). The correspondence between the pictorial approach and the circuit representation used in the IBM Quantum Composer~\cite{IBMcomposer} is shown, and students start practicing with the drag-and-drop interface, creating their own quantum circuits and checking against their pen-and-paper calculations with the pictorial approach. With these tools at hand, the following concepts are taught: quantum key distribution, quantum algorithms (Deutsch, Grover), entanglement (entangled vs separable states), quantum teleportation, and quantum error correction. Finally, after the students spend most of the semester practicing these concepts, the transition to linear algebra is presented by showing how omitting the shapes and keeping coefficients instead allows for states to be represented as vectors. Next, the concept of a matrix as a way to transform the vectors according to the already familiar action of each gate is derived. Most students can actually find the matrices on their own, drawing from their experience and practice with the pictorial representation. 

\section{Syllabus and detailed course description}

\begin{figure*}[htp]
    \includegraphics[width=0.7\textwidth]{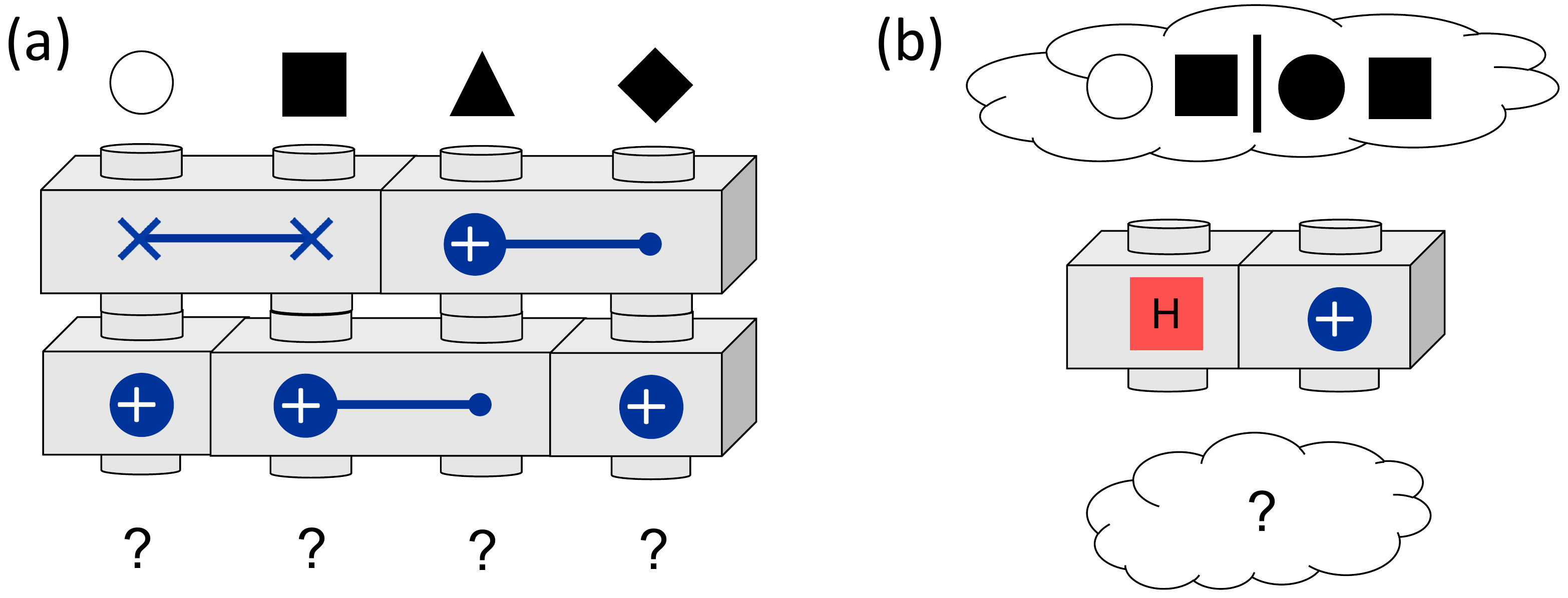}
    \caption{Examples of in-class exercises in which students compute the output of a circuit for a given input state.}
    \label{fig:circuit_problems}
\end{figure*}

\begin{figure*}[htp]
    \includegraphics[width=0.8\textwidth]{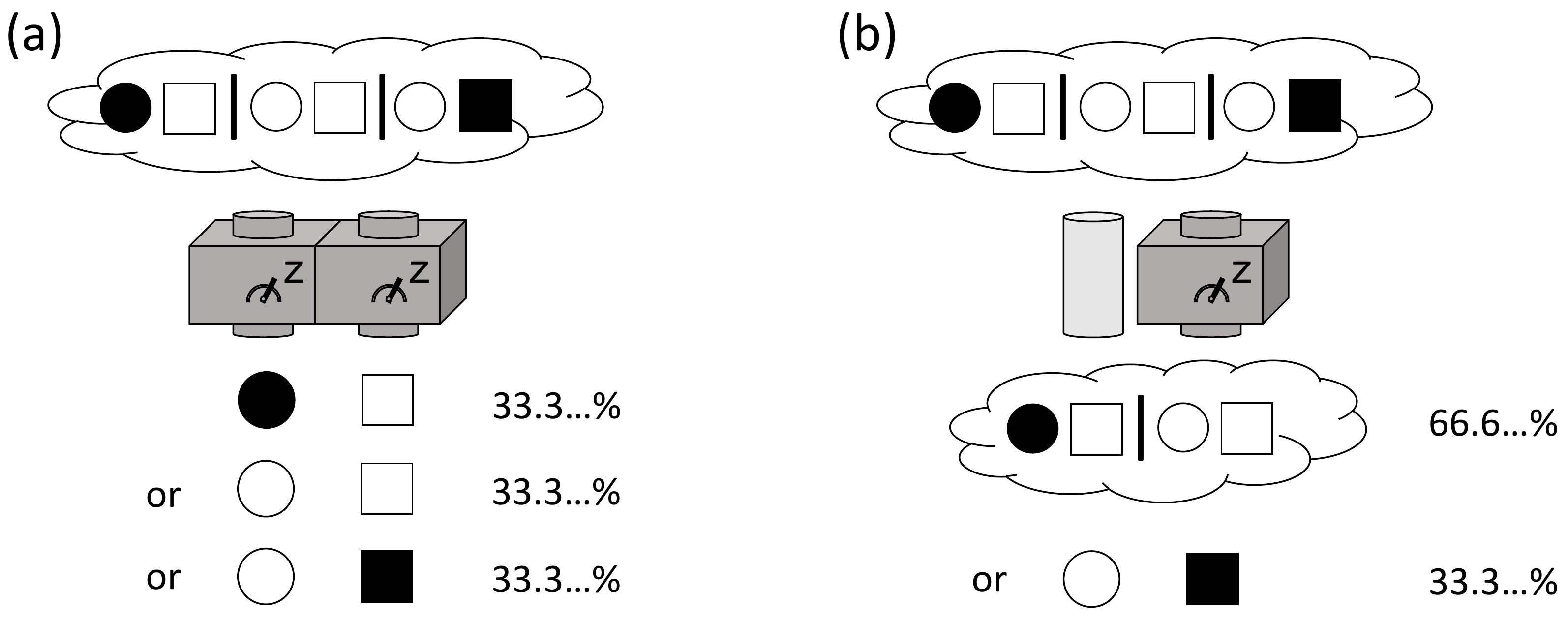}
    \caption{(a) Measuring both qubits of a two-qubit state versus (b) measuring one qubit.}
    \label{fig:measurement}
\end{figure*}

\begin{figure*}[htp]
    \includegraphics[width=0.8\textwidth]{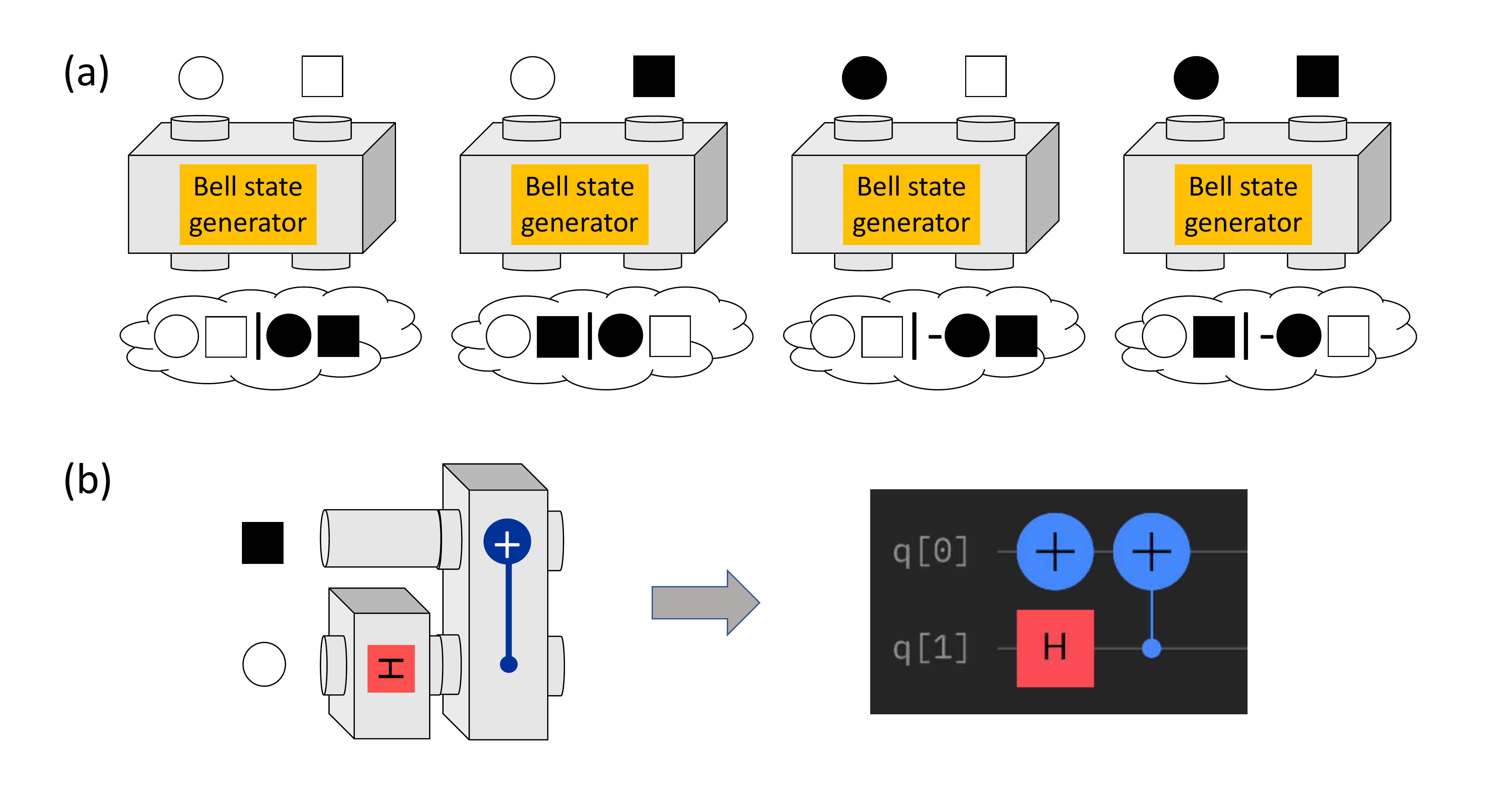}
    \caption{(a) An exercise in which students are asked to find a circuit that is consistent with the given set of input/output rules. (b) An illustration of how a circuit and input state can be mapped to the IBM Quantum Composer by rotating by 90 degrees.}
    \label{fig:entanglement_on_ibm}
\end{figure*}

\begin{figure*}[htp]
    \includegraphics[width=\textwidth]{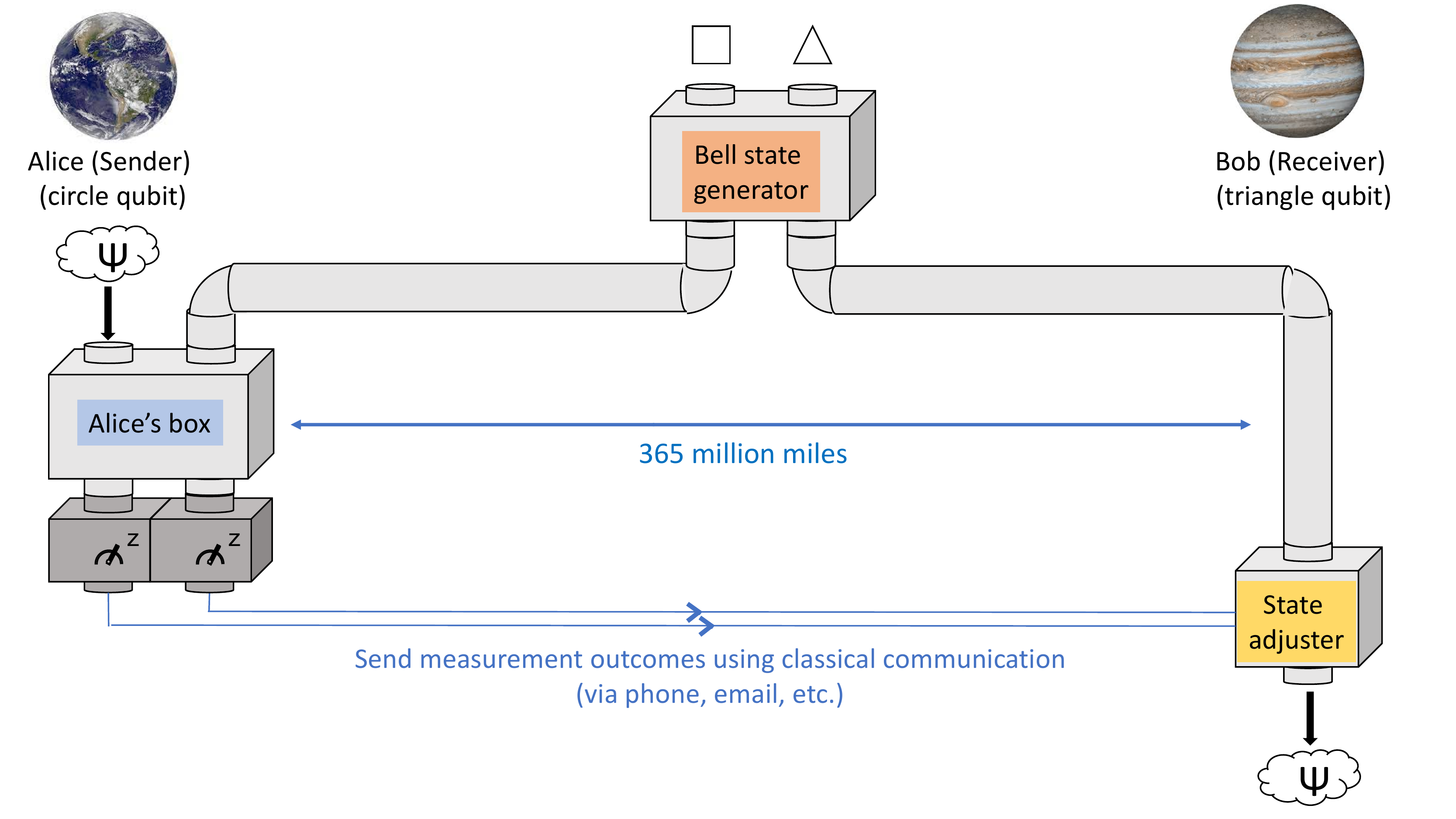}
    \caption{Illustration of the quantum teleportation project that students work on in teams. The students are asked to determine what gates comprise each box shown in the diagram and to program the resulting total circuit on the IBM Quantum Composer for various states $\Psi$.}
    \label{fig:teleportation}
\end{figure*}

\begin{figure*}[htp]
    \includegraphics[width=\textwidth]{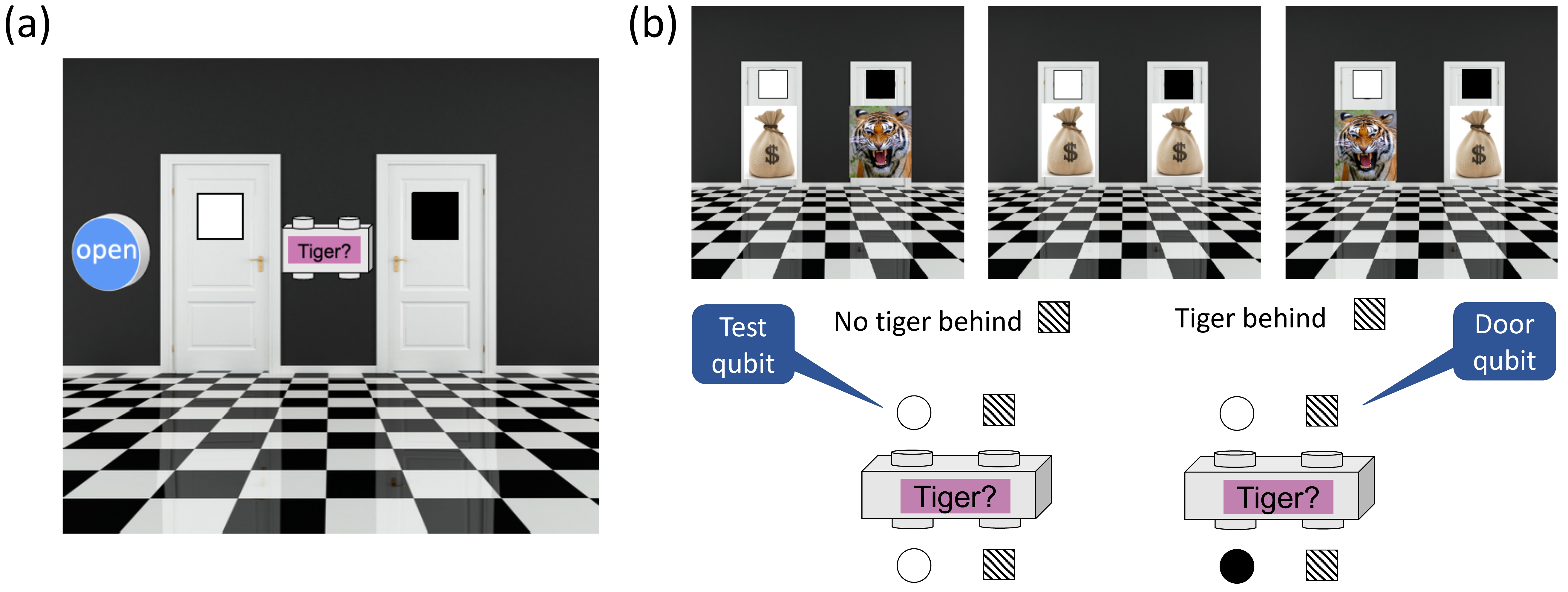}
    \caption{(a) Setup of the Money or Tiger puzzle that students work through in class. There are two doors marked with black and white squares. There is money behind both doors, but there may also be a tiger behind one door. A button on the wall opens both doors at once. A box called Tiger? can be used to check whether there is a tiger behind a given door. (b) The three possible scenarios for where the tiger might be, if there is one. The rules for operating the Tiger? box are also shown. Students figure out how quantum mechanics allows them to use the Tiger? box only once to determine whether the doors should be opened. This is Deutsch's algorithm in disguise. See Ref.~\cite{Economou_arxiv2020} for more details.}
    \label{fig:money_or_tiger}
\end{figure*}

\begin{figure*}
\includegraphics[width=0.6\textwidth]{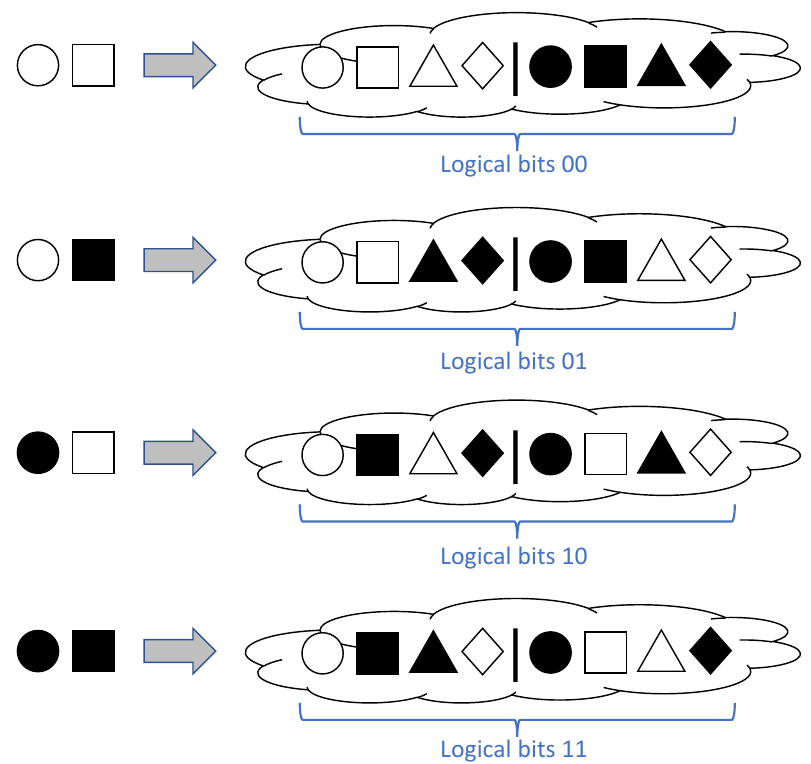}
\caption{The four-qubit error detection code expressed in terms of mists. Students learn about this code in a homework problem, where they confirm the orthogonality of the codewords, work out the logical operations, and construct error-detection circuits.}
\label{fig:error_correction}
\end{figure*}

The course was based on roughly 20 lectures. Below we briefly describe the content of each lecture. 

\begin{itemize}
\item
Lecture 1: Introduction to QISE and overview of the course
\item
Lecture 2: Binary logic, classical bits, classical reversible gates (NOT, CNOT, SWAP, Toffoli), classical circuits (see Fig.~\ref{fig:circuit_problems}(a) for an example in-class problem)
\item
Lecture 3: Double slit experiment, superposition (mist) states
\item
Lecture 4: Hadamard gate, passing superpositions through gates; quantum circuits (see Fig.~\ref{fig:circuit_problems}(b) for an example in-class problem)
\item
Lecture 5: Measurement and probabilities (see Fig.~\ref{fig:measurement})
\item
Lecture 6: Partial measurements; post-measurement states (see Fig.~\ref{fig:measurement})
\item
Lecture 7: Entanglement; entangled versus product states
\item
Lecture 8: Entangled versus product states; circuits for creating entangled states; quantum circuits on IBM Quantum Composer (see Fig.~\ref{fig:entanglement_on_ibm})
\item
Lecture 9: Z and CZ gates
\item
Lecture 10: Bell states and introduction to project on quantum teleportation (see Fig.~\ref{fig:teleportation})
\item
Lecture 11: Orthogonal versus non-orthogonal states; orthogonality criterion; distinguishing orthogonal states; measurement in different bases; measuring in the Bell basis
\item
Lecture 12: No-cloning theorem; quantum key distribution
\item
Lecture 13: Quantum key distribution
\item
Lecture 14: Money or Tiger puzzle (Deutsch algorithm, see Fig.~\ref{fig:money_or_tiger}) and introduction to Grover algorithm project
\item
Lecture 15: Quantum error correction: bit-flip code
\item
Lecture 16: Quantum error correction: phase-flip code; 9-qubit Shor code (the [[4,2,2]] code is given as a HW problem, see Fig.~\ref{fig:error_correction})
\item
Lecture 17: Beyond integer coefficients: transitioning from repeated shapes to coefficients in front of shapes
\item
Lectures 18-20: Transition to linear algebra; mists to vectors; Dirac notation; boxes to matrices
\end{itemize}
During each lecture period, the students work through a variety of practice problems to become familiar with the concepts being introduced. 

Homeworks and projects: Each lecture has an associated set of homework assignments for the students to practice. We also leave time during lectures for students to work on problems hands-on and receive help and feedback in real time. There are two group projects, one on teleportation and the other on `Money or Tigers’. For the former, the students work in teams of three to prepare a presentation on teleportation that addresses a set of questions we provide (see Fig.~\ref{fig:teleportation}). Each team essentially works on their own to discover the concept of teleportation with some prompt questions from us. In the second project, we ask them to again work in groups of three (different teams) to create content, i.e., to make a video that is engaging where they present the `Money or Tigers’ puzzle (Grover's algorithm for four qubits) in any way they choose, so long as it is engaging. Students have come up with very nice and creative ideas.

\section{Discussion}

HQW! was the most enjoyable course either of us has taught to date. The lack of cumbersome math and the concepts-first approach truly engaged, but also challenged, the students. It was very interesting to see that because there was no prerequisite math, and linear algebra was not used, the playing field was level between students of more versus less privileged backgrounds. We believe that oftentimes the students who begin their undergraduate careers better prepared in terms of their mathematical toolbox, and who of course can solve problems faster because they already have knowledge of the math, can appear to be `better students’ or `smarter’ to the instructors, creating a bias early on. By using this entirely new, pictorial tool, all students start from the same place in terms of prior knowledge (for the same reason, using the drag-and-drop IBM Quantum Composer interface for creating circuits, as opposed to writing code with an established programming language, is crucial). At this point, it is important to point out that by doing this, we are not lowering the standards in any way. On the contrary, the material is quite challenging, perhaps more challenging than having the mathematical machinery as a crutch to rely on and blindly solving equations, as sometimes undergraduate students do. The pictorial representation provides an intuitive grasp, not only for the quantum part, but, we have come to believe, also for linear algebra itself. The pictorial approach can be thought of as a distilled abstraction of certain aspects of linear algebra. Concepts such as linearity, basis change, tensor product, have a visual representation the students can refer back to. In fact, students `discover’ the structure of matrices as the inevitable means to describe in a mathematical way the transformations they have been carrying out during the semester. Through in-class hands-on exercises that students solve while we walk around and provide help, we have noticed that the conceptual grasp of these concepts appears to be on average stronger compared to students in our senior-level course on quantum information and computation. Of course, we recognize that this is anecdotal evidence based on a very small sample of students. We hope that this approach will be assessed by professionals in terms of its long-term effectiveness, as we believe there is a lot of opportunity for dramatic improvement of how well students understand quantum information, quantum mechanics, and even linear algebra. 

We have also thought about ways to keep students engaged after the semester finishes. Since we run an annual QISE summer school for high-school students, we recruited some of the undergraduates who successfully completed the HQW! course as teaching assistants for the summer school. This gave them the opportunity to revisit the material almost six months later and to grasp it in a robust way, which is required for them to be able to teach it. We also found that this is a good way for students to feel empowered and continue to be engaged until they, if desired, take on a research project down the road. 

Our excitement about this way of teaching quantum information and computation at the introductory level, along with the interest our course has created among our colleagues, have led us to decide to write a textbook based on this material. Our plan is to make the e-book free to download so that anybody around the world can access it and learn.

\section*{Acknowledgments}

We would like to dedicate this paper to our good friend Terry Rudolph. Terry’s book inspired this journey of ours into teaching quantum information and computation in an accessible way to early college and high-school students, and we have greatly benefited from numerous conversations with him. We also acknowledge support from the National Science Foundation, which over the years has supported our outreach activities; these activities were key in developing and refining the material that lead to Hello Quantum World!. We acknowledge NSF grant nos. 1741656, 1847078, 2137645, 2137953. This material is in part based upon work supported by the U.S. Department of Energy, Office of Science, National Quantum Information Science Research Centers, Co-design Center for Quantum Advantage (C2QA) under contract number DE-SC0012704. We thank Jamie Sikora from the CS Department of Virginia Tech, who has contributed to many of these outreach activities. Finally, we would like to thank the following people from our groups and from groups outside Virginia Tech, mostly from C2QA, for their help with outreach and their feedback on the material: Panagiotis (Peter) Anastasiou, Will Banner, Sam Barron, Yanzhu Chen, Kyle Connelly, Andrew Deutsch, Adrian Florio, Rafail Frantzeskakis, Bryan Gard, Isabela (Bela) Gnasso, Mariano Guerrero Perez, Connor Hann, Zi Huang, Nooshin Mohammadi Estakhri, Ian Mondragon Shem, Hunter Nelson, Richard (Ricky) Oliver, Christian Pederson, Evangelos (Vagelis) Piliouras, Zahra Raissi, Daniel (Danny) Rosen, Karunya Shirali, Kevin Smith, Evangelia (Eva) Takou,  John Van Dyke, Arian Vezvaee, Ada Warren, Linghua Zhu.

%

\end{document}